\documentclass[prd,floatfix,onecolumn,amsmath,amssymb,10pt]{revtex4-2}
\usepackage{newtxtext,newtxmath} 
\usepackage{graphicx,color,dcolumn,booktabs,bm}
\usepackage{subfigure}
\usepackage{amssymb}
\usepackage{longtable}
\usepackage{indentfirst}
\usepackage{epsfig,gensymb,siunitx}
\usepackage{feynmf}   
\usepackage{epstopdf}   
\usepackage{slashed}  
\usepackage{cases}
\usepackage{xcolor}
\definecolor{navyblue}{RGB}{0,0,150}
\definecolor{CobaltBlue}{rgb}{0,0.28,.67}
\definecolor{maroon}{RGB}{139,25,150}
\usepackage{multirow}
\usepackage{float}
\usepackage{pgf}
\usepackage{graphicx,color,dcolumn,booktabs,bm}
\usepackage[colorlinks, citecolor=purple,anchorcolor=purple,menucolor=purple, linkcolor=purple,filecolor=purple,runcolor=purple,urlcolor=purple,frenchlinks=purple, urlcolor=purple]{hyperref}
\usepackage{physics}
\usepackage{orcidlink}
\usepackage{bbold}
\usepackage[utf8]{inputenc} 
\usepackage{varwidth}

\usepackage{xcolor}

\begin{document}
	\preprint{}
\title{\color{navyblue}{Probing the $B^{0}$ meson in a hot medium with finite chemical potential}}
\author{A.~Türkan$^{1}$}
\author{G.~Bozkır$^{2}$}
\author{K.~Azizi$^{3,4}$}
\email{kazem.azizi@ut.ac.ir}
\thanks{Corresponding author}

\affiliation{
	$^{1}$Department of Basic Sciences,  \href{https://hho.msu.edu.tr/pagedetail.aspx?SayfaId=2&ParentMenuId=2}{Air Force Academy, National Defence University}, Yeşilyurt, 34149, İstanbul, Türkiye\\
	$^{2}$Department of Basic Sciences, \href{https://www.msu.edu.tr/} {Army NCO Vocational HE School, National Defence University}, Altıeylül, 10185 Balıkesir,Türkiye\\
	$^{3}$Department of Physics, \href{https://ut.ac.ir/en}{University of Tehran}, North Karegar Avenue, Tehran 14395-547, Iran\\
	$^{4}$Department of Physics, \href{https://www.dogus.edu.tr/en}{Dogus University}, Dudullu-\"{U}mraniye, 34775
	Istanbul,  T\"{u}rkiye}

	\date{\today}

\begin{abstract}

We employ QCD two-point sum rules to determine the $B^{0}$ meson mass and decay constant as functions of temperature $T$ and chemical potential $\mu$, using the perturbative spectral density and nonperturbative contributions up to mass dimension five. Numerical analysis indicates that, in the limits $T \to 0$ and $\mu \to 0$, the mass and decay constant of the $B^{0}$ meson are in good agreement with the available experimental data. Furthermore, at fixed chemical potentials, the $B^{0}$ meson mass and decay constant initially increase with temperature, reach their maxima, and then decrease to zero, with the increase being more pronounced for the decay constant than for the mass. The temperatures at which both quantities vanish decrease as the chemical potential increases. At fixed temperatures, both quantities initially increase slightly with chemical potential before decreasing and eventually vanishing, with the mass exhibiting a much weaker decrease in vacuum. The vanishing points of both quantities shift toward lower chemical potentials as the temperature increases, following a trend similar to that observed in the QCD phase diagram. The results obtained in this study are expected to contribute to the interpretation of data from ongoing heavy-ion collision experiments and to the generation of data for future planned experiments focusing on $B$ mesons, quark–gluon plasma signatures, and dense matter environments.
	
\end{abstract}

\maketitle

\section{Introduction} \label{sec:intro}

The study of pseudoscalar $B$ mesons, consisting of one heavy $b$ quark and one light $u$, $d$, or $s$ quark, plays an important role in the determination of Cabibbo-Kobayashi-Maskawa (CKM) matrix elements in weak interactions and  achieving a better understanding of both perturbative and non-perturbative aspects of QCD. Experimental investigations of the spectroscopic parameters of pseudoscalar $B$ mesons, such as their masses and decay constants, yield essential information on weak interactions and quark mixing, thereby providing valuable insights into the flavor structure of the Standard Model. From the theoretical side, however, the evaluation of these quantities requires the use of non-perturbative approaches, such as lattice QCD, effective field theories, and QCD sum rules, since perturbation theory is not applicable in the low-energy regime of QCD. Using these non-perturbative approaches, the mass and decay constant of pseudoscalar $B$ mesons have been extensively investigated in vacuum~\cite{Neubert, Bazavov1, Bazavov2, Aoki, Black, Albertus, Khan, Lizarazo, Wang, Baker, Lucha, Rahimi, Jamin, Narison, Huang, Bagan,  Ball, Aliev}, while only a few studies have been devoted to their behavior at finite temperature and density~\cite{Chhabra,Chhabra2,Pathak,Kumar1,Mishra}. In addition, their properties in cold nuclear matter have been explored in several works~\cite{Hilger,Wang2,Azizi,Dhale,Arifi,Yasui}. 

Investigation of hadronic properties in a hot and dense medium is important for elucidating the behavior of QCD matter under extreme conditions, such as those produced in relativistic heavy-ion collisions and encountered in the dense interiors of neutron stars. In particular, studying the medium dependence of hadron masses and decay constants at finite temperature and chemical potential may provide valuable information on deconfinement and the restoration of chiral symmetry, and hence contribute to the interpretation of the QCD phase diagram. In this context, while several studies have investigated the in-medium properties of various hadrons in the region characterized by $T>0$ and $\mu>0$ ~\cite{Ayala1,Ayala2,Tawfik,Abu-Shady,Bozkir}, further investigation of heavy-light pseudoscalar mesons containing charm or bottom quarks under such extreme conditions is particularly important given their established role in flavor physics and their sensitivity to medium-induced modifications. In Ref.~\cite{Serna}, the masses of the light pseudoscalar $\pi^+$ and $K^0$ mesons and the heavy-light pseudoscalar $D^+$ meson were investigated at finite temperature and chemical potential using a Dyson–Schwinger/Bethe–Salpeter approach with a vector–vector contact interaction. Their numerical results show that the $D^+$ meson mass remains approximately constant at low temperatures and increases at higher temperatures, with this increase beginning at progressively lower temperatures as the chemical potential increases. In Ref.~\cite{Blaschke}, the in-medium properties of the heavy-light pseudoscalar $D^+$ and $D^-$ mesons  were investigated at finite temperature and chemical potential within the Nambu–Jona-Lasinio (NJL) and Polyakov–Nambu–Jona-Lasinio (PNJL) models. The authors found that Pauli blocking induces a moderate mass splitting of up to 100 MeV between the $D^+$ and $D^-$ mesons, with the $D^-$ mass increasing while the $D^+$ mass remains approximately constant. Beyond these studies on charmed mesons, investigating the in-medium properties of their bottom counterparts in hot and dense matter is also of interest, since differences in their heavy-quark masses may lead to different responses to the medium. The effects of finite temperature and density on the masses and decay constants of the pseudoscalar $D$, $B$, $D_s$, and $B_s$ mesons have also been investigated using QCD sum rules and the chiral $SU(3)$ model~\cite{Chhabra,Chhabra2}. The authors reported that increasing the density leads to larger negative mass and decay constant shifts for these mesons at fixed temperature, isospin asymmetry, and strangeness fraction. At finite density, increasing the temperature from $T=0$ to $T=100~\mathrm{MeV}$ was found to reduce the magnitude of these negative shifts. Isospin asymmetry was also found to induce mass splitting between the charged and neutral $D$ and $B$ mesons~\cite{Chhabra}. For the $D_s$ and $B_s$ mesons, larger relative reductions were reported for the $B_s$ meson, particularly in the decay constant~\cite{Chhabra2}. In the same work, the authors also calculated the in-medium partial decay widths of the excited $D_s^{*}(2715)$ and $D_s^{*}(2860)$ states into the $D_s(1968)\eta$ channel within the $^3P_0$ model by incorporating the medium modifications of the final state meson masses. In Ref.~\cite{Pathak}, the in-medium behavior of $B$ and $\bar{B}$ mesons has been examined in hot, dense, isospin-asymmetric strange hadronic matter using a chiral effective model. The authors showed that both $B$ and $\bar{B}$ mesons experience attractive interactions in the medium, leading to a reduction in their masses. They also found that the medium breaks the particle–antiparticle mass degeneracy, while isospin asymmetry induces mass splitting within the isospin doublets. Moreover, the medium effects were reported to depend strongly on density and only weakly on temperature. In Ref.~\cite{Kumar1}, the effects of finite temperature and density on the masses and scattering lengths of the scalar $D_0$ and $B_0$, vector $D^{\ast}$ and $B^{\ast}$, and axial-vector $D_1$ and $B_1$ mesons were analyzed within QCD sum rules using medium-modified quark and gluon condensates obtained from the chiral $SU(3)$ model. The analysis showed that, at finite temperature and density, the scalar and axial-vector mesons acquire positive mass shifts relative to their vacuum values, whereas the vector mesons exhibit negative mass shifts. At fixed density, both the positive mass shifts and the magnitudes of the negative mass shifts decrease with increasing temperature. In addition, the magnitudes of the scattering lengths of the scalar, vector, and axial-vector mesons were found to decrease with increasing density at fixed temperature and with increasing temperature at fixed density. In Ref.~\cite{Mishra}, the in-medium masses of the pseudoscalar $D$, $\bar{D}$, $D_s$, $\bar{D}_s$, $B$, $\bar{B}$, $B_s$, and $\bar{B}_s$ mesons were investigated at finite temperature and density in isospin-asymmetric strange hadronic matter within the framework of QCD sum rules, with the medium-dependent quark and gluon condensates obtained from the chiral $SU(3)$ model. The authors found that the masses of the particles $D^+$ ($D^0$) and $B^-$ ($\bar{B}^0$) in nuclear matter were smaller than those of their corresponding antiparticles $D^-$ ($\bar{D}^0$) and $B^+$ ($B^0$). In isospin-symmetric matter, they showed that both the non-strange particles and their antiparticles exhibit positive mass shifts relative to their vacuum values, with larger shifts for the antiparticles. Their numerical results indicate that these positive mass shifts generally increase with density.

As discussed above, medium modifications of $B$ mesons at finite temperature and density have mainly been investigated using chiral effective models or QCD sum rules combined with the chiral $SU(3)$ model~\cite{Chhabra,Chhabra2,Pathak,Kumar1,Mishra}. However, the dependence of the $B^{0}$ meson mass and decay constant on temperature and chemical potential remains relatively unexplored within a QCD sum rule framework incorporating a spectral density and condensates that depend on both $T$ and $\mu$.  Investigating these spectroscopic parameters in the $T$–$\mu$ plane, which is commonly used to represent the QCD phase diagram, may provide further insight into the behavior of bottom mesons in hot and dense QCD matter. Therefore, in the present work, we investigate these spectroscopic parameters at finite temperature and chemical potential within the framework of QCD sum rules (QCDSR).  For this purpose, we first obtain the perturbative spectral density as a function of temperature $T$ and chemical potential $\mu$. We then incorporate the non-perturbative contributions that depend on both parameters, including operators up to mass dimension five. By matching the QCD and physical representations of the correlation function, we derive the sum rules for the mass and decay constant of the $B^{0}$ meson in a hot and dense medium. After establishing suitable working intervals for the auxiliary parameters, we numerically analyze the resulting sum rules and examine the dependence of the mass and decay constant on temperature and chemical potential.

The rest of this paper is organized in the following way: In Section \ref{sec:1}, the QCDSR formalism for the mass and decay constant of the $B^{0}$ meson at finite temperature and chemical potential is constructed, with separate discussions of the QCD and physical frameworks. Section \ref{sec:2} is devoted to the numerical analysis of the resulting sum rules and the behavior of these spectroscopic parameters in a hot medium with finite chemical potential. In Section \ref{sec:3}, the main results are summarized and concluding remarks are presented. 

\section{QCDSR FORMALISM IN A HOT MEDIUM WITH FINITE CHEMICAL POTENTIAL} \label{sec:1}

One of the most fundamental and challenging questions in particle physics is to identify the critical point that separates the hadronic and quark–gluon plasma phases in the QCD phase diagram. A substantial body of experimental and theoretical work has been devoted to addressing this question; nevertheless, a commonly accepted conclusion has yet to emerge, keeping the subject at the forefront of ongoing research. Within this framework, among the non-perturbative approaches to hadron phenomenology, QCDSR constitute a powerful theoretical tool for exploring hadronic properties by linking observable hadronic quantities to the underlying quark–gluon dynamics of QCD.

 The foundations of the QCDSR were originally developed by Shifman, Vainshtein and Zakharov to explore the vacuum properties of mesons  \cite{Shifman1,Shifman2} and were subsequently extended to the baryonic sector by Ioffe \cite{Ioffe}. Its formulation was later generalized to finite-temperature systems by Bochkarev and Shaposhnikov \cite{Bochkarev}, giving rise to thermal version of QCDSR. In this extension, the vacuum formulation is modified by replacing vacuum condensates with their thermal counterparts and by incorporating additional operators into the operator product expansion that reflect the presence of the medium and its preferred four-velocity.

In the QCDSR framework, the determination of the mass and decay constant plays a fundamental role in the investigation of other hadronic properties. In particular, the temperature and chemical potential dependence of the decay constant provides an essential probe of chiral symmetry restoration, deconfinement, and hadron melting, while its vacuum value plays a crucial role in the extraction of the CKM matrix elements. Accordingly, the time-ordered two-point correlation function constitutes the starting point for the determination of the $B^0$ meson mass and decay constant in QCDSR and is defined as
\textbf{}
\begin{equation}
	\label{correlation}
	\Pi(T,\mu) = i \int d^4 x e^{ip\cdot x} \langle\psi|\mathcal{T}[{J}_{B^{0}}(x){J}_{B^{0}}^\dagger(0)]|\psi\rangle,
\end{equation}
where  $p=(p_{0},\textbf{p}) $  is the four-momentum of the particle, $|\psi\rangle$ represents the ground state of the hot medium with finite chemical potential. In the vacuum limit, $|\psi\rangle$ reduces to the vacuum state $|0\rangle$. The symbol $\mathcal{T}$ stands for the time-ordering operator, while the interpolating current of the $B^{0}$ meson is given by
\begin{eqnarray}\label{QCD}
	{J}_{B^{0}}(x)=\bar{b}^a(x)i\gamma_{5} d^a(x).
\end{eqnarray} 
Here $b(x)$ and $d(x)$ represent the bottom and down quark fields,  respectively and $ \gamma_{5}$ is the standard Dirac matrix. 
In the context of QCD sum rules, the correlation function given in Eq. (\ref{correlation}) can be calculated in two different momentum regions. This allows us to obtain information on both the hadronic properties of the particle under investigation and its internal structure, which is described in terms of the underlying quark fields and quantum numbers.

\subsection{QCD framework: spectral density and non-perturbative contributions} 

In the deep Euclidean region, the correlation function is evaluated with the help of the operator product expansion (OPE), which enables us to separate the perturbative and non-perturbative contributions. This expression is referred to as the QCD side of the correlator and is written as
\begin{eqnarray}\label{QCD}
	\Pi^{QCD}(T,\mu)&=&\Pi^{pert.}(T,\mu)+\Pi^{np.}(T,\mu)\nonumber \\
	&=&C_{0}\textbf{I}+\sum_{d}C_{d}(x)\langle\psi|\hat{O}_{d}(0)|\psi\rangle.
\end{eqnarray} 
In Eq. (\ref{QCD}),  $C_{d}(x)$ stands for the Wilson coefficients and $\hat{O}_{d}$ represent local operators with increasing mass dimension $  d$.  The first term on the right-hand side corresponds to the perturbative contribution associated with the lowest-dimensional operator $  d=0$ and $ \textbf{I} $ denotes the unit matrix. The remaining operators in the expansion encode the non-perturbative effects and the lowest-dimensional color-singlet non-perturbative operator is the quark condensate $  \langle \bar{q}q \rangle$ with mass dimension $ d=3$. In this calculation, the contributions of operators up to mass dimension five are taken into consideration.

To derive the perturbative part of the correlator, the dispersion integral is defined in terms of the spectral density, $ \rho(T,\mu) $, as
\begin{equation}\label{dispersion}
	\Pi^{QCD}(T,\mu)=\int ds\frac{\rho(T,\mu)}{s-p^{2}}+\Pi^{np.}(T,\mu),
\end{equation}
where the temperature dependence of the spectral density at finite chemical potential is given by
\begin{equation}\label{spectral}
	\rho(T,\mu)=\frac{1}{\pi}Im \Pi^{pert.}(T,\mu) \tanh \Big(\beta\frac{p_{0}}{2}\Big).
\end{equation}
Here $ \beta $ is the inverse temperature, $ \beta=1/T$. In momentum space, for the region $ p^{2}\ll-\Lambda^{2}_{QCD} $, the  amplitude corresponding to the bare loop is written as
\begin{equation}\label{bareloop}
	\Pi^{pert.}(T,\mu)=i N_{c}\int \frac{d^{4}q}{(2 \pi)^{4}} Tr \Big[ S(q)i\gamma_{5} S(p+q)i\gamma_{5} \Big],
\end{equation}
where $ N_{c} = 3 $ is the number of colors and the in-medium quark propagator takes the form \cite{tez}
\begin{equation}\label{propagator}
	S(p)=\frac{i(\!\not\!{p}+m)}{p^{2}-m^{2}+i\varepsilon} -2\pi n_{f} (| p_{0}|)(\!\not\!{p}+m) \delta(p^{2}-m^{2}).
\end{equation}
In medium-modified quark propagator, the temperature dependence is introduced through the Fermi-Dirac distribution function, $n_{f}(x)$, defined as $n_{f}(x)=1/(e^{\beta x}+1)$, whereas  the chemical potential dependence is incorporated through the zeroth component of the four-momentum, $\acute{p_0}=p_{0}+\mu$ with $p_0=\sqrt{s}$. Substituting the propagator given in Eq. (\ref{propagator}) into Eq. (\ref{bareloop}), we obtain
\begin{eqnarray}\label{bareloop2}
	\Pi^{pert.}(T,\mu)&=&3i\int \frac{d^{3}\textbf{q}}{(2 \pi)^{4}}\int dq_{0}Tr \Big((\!\not\!{q}+m_{d})\gamma_{5} (\!\not\!{p}+\!\not\!{q}+m_{b})\gamma_{5} \Big)
	\Big[\frac{1}{q^{2}-m_{d}^{2}+i\varepsilon} \frac{1}{(p+q)^{2}-m_{b}^{2}+i\varepsilon}\nonumber \\
	&+&\frac{1}{q^{2}-m_{d}^{2}+i\varepsilon} 2\pi i n_{f}(| p_{0}+\mu+q_{0} |) \delta((p+q)^{2}-m_{b}^{2})+\frac{1}{(p+q)^{2}-m_{b}^{2}+i\varepsilon} 2\pi i n_{f}(| q_{0} |) \delta(q^{2}-m_{d}^{2})\nonumber \\
	&-&4\pi^{2} n_{f}(|q_{0}|) n_{f}(|p_{0}+\mu+q_{0}|)\delta(q^{2}-m_{d}^{2})
	\delta((p+q)^{2}-m_{b}^{2})\Big].
\end{eqnarray}
By evaluating the Feynman integrals using the Cutkosky rule, $ (q^{2}-m^{2}+i\varepsilon)^{-1}=(-2i \pi)\delta(q^{2}-m^{2})$, in Eq. (\ref{bareloop2}), the following expression is ultimately obtained. The calculations are performed in the rest frame of the initial particle, $\textbf{p}=0 $:
 \begin{eqnarray}\label{QCDpert}
	\Pi^{pert.}=\frac{3i(\sqrt{s}+\mu)^{2}-(m_{b}-m_{d})^{2})^{2}}{ 8\pi^{2}(\sqrt{s}+\mu)^{2}}\sqrt{1-\frac{4m_{d}m_{b}}{(\sqrt{s}+\mu)^{2}-(m_{b}-m_{d})^{2}}}(1-n_{f}(|\omega_{1}|)-n_{f}(|\omega_{2}|)+2n_{f}(|\omega_{1}|)n_{f}(|\omega_{2}|)),
\end{eqnarray}
with
\begin{eqnarray}\label{FermiDirac}
	&&n_{f}(|\omega_{1}|)=n_{f}(|\frac{m_{d}^{2}-m_{b}^{2}+(\sqrt{s}+\mu)^{2}}{2(\sqrt{s}+\mu)}|),\nonumber \\ 
	&&n_{f}(|\omega_{2}|)=n_{f}(|(\sqrt{s}+\mu)-\omega_{1}|).
\end{eqnarray}
The final form of the spectral density is obtained by using the relation
given in  Eq. (\ref{spectral}) and resulting in the following expression  
\begin{eqnarray}\label{SD}
	\rho(T,\mu)&=&\frac{3((\sqrt{s}+\mu)^{2}-(m_{b}-m_{d})^{2})^{2}}{8\pi^{2} (\sqrt{s}+\mu)^{2}}\sqrt{1-\frac{4m_{d}m_{b}}{(\sqrt{s}+\mu)^{2}-(m_{b}-m_{d})^{2}}}(1-n_{f}(|\omega_{1}|)-n_{f}(|\omega_{2}|)).
\end{eqnarray}
The spectral density obtained in Eq.  (\ref{SD})  corresponds to the Wilson coefficient $C_0$ associated with the perturbative contribution. The non-perturbative part of the correlator can then be expressed in terms of the quark condensate $\langle\bar{d}d  \rangle_{T,\mu}$,  the gluon condensate $\langle G_{\mu \nu}G^{\mu \nu} \rangle_{T,\mu}=\langle G^2 \rangle_{T,\mu}$ where $G_{\mu\nu}$ is the gluon field and the mixed condensate $ \langle\bar{d}g_{s}\sigma_{\mu \nu} G^{\mu \nu}d\rangle_{T,\mu}$ with $ \sigma_{\mu \nu} =\frac{i}{2}[\gamma_\mu, \gamma_\nu]$ as follows

\begin{eqnarray}\label{OPE}
	\Pi^{np.}(T,\mu)&=&C_{3} \langle\bar{d}d  \rangle_{T,\mu}+C_{4a} \langle G^2 \rangle_{T,\mu}+C_{4b} \langle \Theta_{\mu\nu} \rangle_T +C_5  \langle\bar{d}g_{s}\sigma Gd\rangle_{T,\mu}.
\end{eqnarray} 
In Eq. (\ref{OPE}), $ \langle \Theta_{\mu\nu} \rangle$ is energy momentum tensor with mass dimension four. In order to obtain the Wilson coefficient of condensates ($C_3, C_{4a},C_{4b},  C_5$) the following expression must be computed.
\begin{eqnarray}\label{nonpert}
	\Pi^{np.}(T,\mu)&=&\langle\psi|\bar{d}_{\alpha}^{i}(0)\Big(i\gamma_{5}\frac{i}{\!\not\!{p}-m_b}i\gamma_{5}\Big)_{\alpha\beta}d_{\beta}^{j}(x)|\psi\rangle\nonumber \\
	&+&\int \frac{d^{4}q}{(2 \pi)^{4}} \int \frac{d^{4}k_1}{(2 \pi)^{4}}\int \frac{d^{4}k_2}{(2 \pi)^{4}} \nonumber \\
	&\times& \Big\{ Tr \Big[ i\gamma_{5} \frac{i}{\!\not\!{p}+\!\not\!{q}-\!\not\!{k_1}-\!\not\!{k_2}-m_b}\Gamma_2\frac{i}{\!\not\!{p}+\!\not\!{q}-\!\not\!{k_1}-m_b}\Gamma_1 \frac{i}{\!\not\!{p}+\!\not\!{q}-m_b} i\gamma_{5}\frac{i}{\!\not\!{q}-m_d} \Big]  \nonumber \\
	&+&Tr \Big[ i\gamma_{5} \frac{i}{\!\not\!{p}+\!\not\!{q}-\!\not\!{k_2}-m_b}\Gamma_2\frac{i}{\!\not\!{p}+\!\not\!{q}-m_b}i\gamma_{5}\frac{i}{\!\not\!{q}-m_d}\Gamma_1  \frac{i}{\!\not\!{q}+\!\not\!{k_1}-m_d} \Big]  \nonumber \\
	&+&Tr \Big[ i\gamma_{5} \frac{i}{\!\not\!{p}+\!\not\!{q}-m_b}i\gamma_{5}\frac{i}{\!\not\!{q}-m_d}\Gamma_1 \frac{i}{\!\not\!{q}+\!\not\!{k_1}-m_d}\Gamma_2 \frac{i}{\!\not\!{q}+\!\not\!{k_1}+\!\not\!{k_2}-m_d} \Big]\Big \} \nonumber \\
	&+&\langle\psi|\bar{d}_{\alpha}^{i}(0)\Big(i\gamma_{5}\frac{i}{\!\not\!{p}-m_b}\Gamma \frac{i}{\!\not\!{p}+\!\not\!{k}-m_b}i\gamma_{5}\Big)_{\alpha\beta}d_{\beta}^{j}(x)|\psi\rangle ,
\end{eqnarray}
where $ \Gamma_{(1,2)}=i\frac{g_s}{2}x_{\lambda}\gamma_{\tau}G_{\lambda\tau} $ , $g_s$ denotes strong coupling and $ x_{\lambda}=-i \frac{d}{d_\lambda} $. To evaluate the expressions given above, the quark field defined at point $x$ is replaced by its Taylor series expansion around $x=0$ as follows
\begin{equation}\label{taylor}
	d_{\beta}(x)=d_{\beta}(0)+x_{\alpha}\nabla_{\alpha}d_{\beta}(0)+\frac{1}{2}x_{\alpha}x_{\alpha'}\nabla_{\alpha}\nabla_{\alpha'}d_{\beta}(0)+~\cdots.
\end{equation}

To carry out the subsequent calculations, in addition to the Taylor expansion of the quark fields, the traces of the thermal expectation value of the product of two gluon field-strength in the medium must also be specified. The required relation is given in terms of the medium-dependent gluon condensate $ \langle G^2\rangle_{T,\mu} $ and the additional operators associated with the gluonic part of energy momentum tensor $\langle \Theta^{g}_{\mu\nu} \rangle$ \cite{Mallik} as
\begin{eqnarray}\label{TrGG} 
	\langle Tr^c G_{\alpha \beta} G_{\mu \nu}\rangle_{T,\mu} &=& \frac{1}{24} (g_{\alpha \mu} g_{\beta \nu} -g_{\alpha
		\nu} g_{\beta \mu})\langle G^2\rangle_{T,\mu} \nonumber \\
		&+&\frac{1}{6}\Big[g_{\alpha \mu}g_{\beta \nu} -g_{\alpha \nu} g_{\beta \mu} -2(u_{\alpha} u_{\mu}g_{\beta \nu} 
	-u_{\alpha} u_{\nu} g_{\beta \mu} -u_{\beta} u_{\mu}
	g_{\alpha \nu} +u_{\beta} u_{\nu} g_{\alpha \mu})\Big]\langle u^{\lambda} {\Theta}^g _{\lambda \sigma} u^{\sigma}\rangle.
\end{eqnarray}
Here $u^{\mu}=(1,0,0,0)$ is the four-velocity vector, which is introduced to account for the breaking of Lorentz invariance resulting from the choice of the thermal rest frame in the Wilson expansion. After lengthy but straightforward calculations, the Wilson coefficients corresponding to the non-perturbative contributions in the  Borel scheme ($ \hat{\textbf{B}} $) are given as follows
\begin{eqnarray}\label{}
	&& \hat{\textbf{B}} C_3= \Big( \frac{2m_{b}+m_{d}}{2}+\frac{m_{b}^{2}m_{d}(m_{b}m_{d}+M^2)}{2M^{4}} \Big) e^{-m_b^2/M^2},\nonumber \\ 
	&& \hat{\textbf{B}} C_{4a}= \Big( \frac{m_b^3m_d(z-1)-(12m_b^2-2m_bm_d)M^2(z-1)^2-12M^4(z-1)^3+m_b^4z}{24M^4(z-1)^3} \Big) e^{-m_b^2/M^2(z-1)},\nonumber \\ 
	&& \hat{\textbf{B}} C_{4b}= \Big( -\frac{g_s^2(T)(m_b^2+(M^2+4p_0^2)(z-1))}{6M^2\pi^2(z-1)}e^{-m_b^2/M^2(z-1)}, \nonumber \\ 
	&& 	\hat{\textbf{B}}C_{5}= \Big( -\frac{m_b(m_b^2-2M^2)}{4M^4} \Big) e^{-m_b^2/M^2}.
\end{eqnarray}
where $M^2$ is the Borel mass parameter.
\subsection{Physical framework: hadron mass and decay constant}

In the second momentum region, known as the physical side, the same correlation function can be evaluated by writing it in terms of hadronic parameters, including the mass and decay constant of the $B^{0}$ meson. The physical side of the correlator given in Eq. (\ref{correlation}) is obtained by inserting complete sets of intermediate hadronic states between the interpolating currents and can be written as
 \begin{equation}\label{physical}
 	\Pi^{Phys.}(T,\mu)  =\frac{ \langle\Psi|J_{B^{0}}|B^{0}\rangle  \langle B^{0}|J^\dagger_{B^{0}}|\Psi \rangle}{m_{B^{0}}^{*2}-p^{2}}  +~\cdots,
 \end{equation}
 where the ellipsis $\cdots$ indicates the contributions of the higher and continuum states for corresponding particle. The matrix element can be parameterized in terms of the $B^{0}$ meson mass and decay constant as
\begin{eqnarray}\label{matrix}
\langle \Psi|J_{B^{0}}|B^{0}\rangle= i \frac{m_{B^{0}}^{*2}f_{B^{0}}^{*}}{m_b+m_d}.
\end{eqnarray}
For brevity, the temperature and chemical potential dependent quantities $m_{B^{0}}(T,\mu)$ and $f_{B^{0}}(T,\mu)$ are denoted by $m^{*}$ and $f^{*}$, respectively. Substituting Eq. (\ref{matrix})  into Eq.  (\ref{physical}) and subsequently applying the Borel transformation, the final form of the physical side is obtained as follows
\begin{eqnarray}\label{physicalborel}
		\hat{\textbf{B}}\Pi^{Phys.}(T,\mu) &=& -\frac{m_{B^{0}}^{*4}f_{B^{0}}^{*2}}{(m_b+m_d)^2}e^{-m_{B^{0}}^{*2}/M^{2}} \textbf{I}.
\end{eqnarray}
\subsection{Mass and decay constant sum rules}
To obtain the sum rules for the mass and decay constant of the particle under investigation, the final step is to match the physical and QCD representations of the correlation function, {obtained in two different momentum regions $(\hat{\textbf{B}}\Pi^{Phys.}(T,\mu) =\hat{\textbf{B}}\Pi^{QCD}(T,\mu))$. This procedure requires applying continuum subtraction as well as the Borel transformation in order to suppress the contributions of higher resonances and continuum states. Finally we find the desired sum rules
\begin{eqnarray}\label{sumrule}
\frac{m_{B^{0}}^{*4}f_{B^{0}}^{*2}}{(m_b+m_d)^2}e^{-m_{B^{0}}^{*2}/M^{2}}&=&\int_{(m_{b}+m_{d})^2}^{s_{0}(T,\mu)}ds\rho(T,\mu)e^{-s/M^2}+\hat{\textbf{B}} \Pi^{np.}(T,\mu).
\end{eqnarray}
Using Eq. (\ref{sumrule}) the temperature and chemical potential dependent sum rules for the mass and decay constant are extracted as
\begin{eqnarray}\label{massSM}
	m_{B^0}^{*2}=\frac{\frac{d}{d(-1/M^2)} \Big [\int_{(m_{b}+m_{d})^2}^{s_{0}(T,\mu)}ds\rho(T,\mu)e^{-s/M^2}+\hat{\textbf{B}} \Pi^{np.}(T,\mu) \Big ] } {\int_{(m_{b}+m_{d})^2}^{s_{0}(T,\mu)}ds\rho(T,\mu)e^{-s/M^2}+\hat{\textbf{B}} \Pi^{np.}(T,\mu)},
\end{eqnarray}
\begin{eqnarray}\label{decaySR}
	f_{B^0}^{*2}=\frac{(m_b+m_d)^2}{m_{B^0}^{*4}}  \Big [\int_{(m_{b}+m_{d})^2}^{s_{0}(T,\mu)}ds\rho(T,\mu)e^{-s/M^2}+\hat{\textbf{B}} \Pi^{np.}(T,\mu) \Big ]  e^{m_{B^0}^{*2}/M^2}.
\end{eqnarray}

\section{Numerical Analysis of QCDSR}  \label{sec:2}

In this section, we carry out a detailed numerical analysis of the derived sum rules to examine the dependence of the mass and decay constant of the $B^{0}$ meson on temperature and chemical potential, and to obtain the corresponding vacuum values in the limit $T=\mu=0$. To perform this analysis, a set of appropriate input parameters must be specified, including the quark masses and the QCD condensates relevant to both the vacuum and the medium. 
\begin{itemize}
	\item \textbf{Behavior at zero temperature and zero chemical potential:}
Since the medium-dependent condensates are constructed from their vacuum counterparts, we first specify the corresponding vacuum values of the quark, gluon, and mixed condensates. The numerical values adopted in our analysis are summarized in Table \ref{tab:Param}.
\begin{table}[ht!] 
	\centering
	\begin{tabular}{ |c|c|c|c|}
		\hline \hline 
		Parameter  &  Value & Unit & Reference  \\ \hline
		$q_0 $  &  $ 5279.63 \pm0.20 $   & $MeV$& \cite{PDG}  \\   
		$ m_{u}   $    &  $2.16_\pm0.07$  & $MeV$& \cite{PDG} \\   
		$ m_{d}   $       & $4.70\pm0.07$  &$MeV$ & \cite{PDG}  \\
		$  m_{s}   $          &  $93.4_{-3.4}^{+8.6} $    & $MeV$& \cite{PDG}   \\
		$ m_{b}   $         &   $4.186\pm0.006$  & $GeV$&  \cite{PDG}\\
		$ \langle \overline{q}q \rangle _{0,0} (1~GeV)$          &  $-(0.24 \pm 0.01)^3$    & $GeV^3$ &\cite{Belyaev}   \\
		$ {\langle}  \frac{\alpha_s}{\pi} G^2 {\rangle}_{0,0}$          &  $ 0.012 \pm 0.004$ & $GeV^4$ & \cite{Belyaev1} \\
		$\langle 0| \bar{q}g_{s}\sigma_{\mu \nu} G^{\mu \nu} q|0 \rangle$         &   $m_{0}^{2} \langle  \bar{q}q \rangle_{0,0} $ &$GeV^5$ & \cite{Belyaev}\\
		$m_{0}^{2};   $          &  $0.8\pm0.2$   & $GeV^2$& \cite{Belyaev}  \\
		
		\hline \hline
	\end{tabular}
	\caption{Vacuum input parameters adopted in the analysis.}
	\label{tab:Param}
\end{table}

Prior to incorporating the temperature and chemical potential dependence of the condensates, it is necessary to establish the admissible ranges of the two adjustable parameters appearing in the sum rule relation formulated through Eq.~(\ref{sumrule}), namely the Borel mass parameter, $M^2$, and the vacuum continuum threshold, $s_0(0,0)$. Their allowed ranges are constrained by imposing the conventional QCD sum rule requirements, thereby ensuring the robustness and reliability of the numerical analysis. In particular, the upper bound of $M^2$ is chosen such that the pole contribution associated with the ground state remains larger than the combined contribution of the continuum and higher resonances.  Conversely, the lower bound is established by requiring a well-convergent OPE, where the perturbative term provides the dominant contribution and successive condensate corrections exhibit a convergent behavior. At zero temperature and chemical potential, the continuum threshold $s_0(0,0)$ is chosen with reference to the energy region associated with the lowest excited state and is therefore not treated as an arbitrary parameter. In addition to this spectral constraint, the admissible range of $s_0(0,0)$ is determined by requiring both a sufficiently dominant pole contribution and a stable OPE convergence. Taking these conditions into account, the resulting working intervals are given by
\begin{equation}\label{eqs1}
	M^2 \in [8 - 10]~\text{GeV}^2~~~~~~, \qquad s_0 (0,0)\in [30 - 32]~\text{GeV}^2.
\end{equation}
To assess the stability of the extracted vacuum mass against variations in  $M^2$ and  $s_0(0,0)$ within their respective working regions, we display, as an illustrative example, the dependence of the mass on these two parameters in a three-dimensional plot shown in Fig. (\ref{vacuumMassG}). For the construction of Fig. (\ref{vacuumMassG}), medium modified condensates were considered in the limits $T\rightarrow 0$ and $\mu \rightarrow 0$; their explicit  expressions and definitions are provided in the following subsection. The plot indicates that the extracted mass exhibits only a weak sensitivity to variations in both $M^2$ and $s_0(0,0)$, with the resulting changes remaining within the acceptable limits of the sum rule analysis.
\begin{figure}[!h]
	\centering
	\begin{minipage}{0.48\textwidth}
		\centering
		\includegraphics[width=\linewidth]{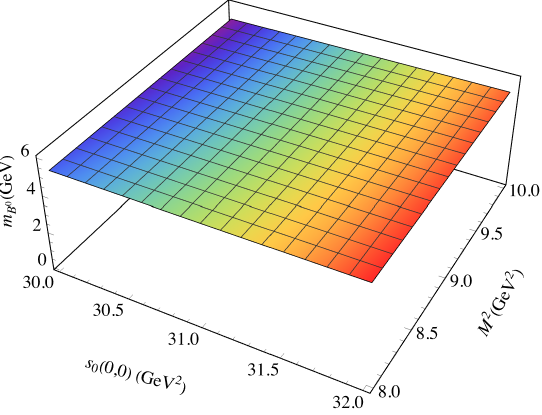}
		\label{fig:left}
	\end{minipage}
	\vspace{-10pt}
	\caption{Variation of the $ B^0$ meson mass versus $M^2$ and $ s_{0} (0,0)$ at zero temperature and chemical potential.}\label{vacuumMassG}
\end{figure}
The reliability of the derived medium-dependent sum rules can be assessed by examining whether they reproduce the experimentally measured values of the quantities under study in the vacuum limit. Using the  values of the Borel parameter and continuum threshold, the corresponding vacuum mass and decay constant are then obtained as follows
\begin{eqnarray}\label{zeromassandf}
	&& m_{B^0} (0,0)= 5.297_{-0.039}^{+0.031}  \, \textrm{GeV},  \nonumber \\ 
	&&  f_{B^0}(0,0) =0.189_{-0.025}^{+0.056}  \, \textrm{GeV}.
\end{eqnarray}
The uncertainties quoted above originate from the parameter windows specified in Eq. (\ref{eqs1}) as well as other inputs.  Our results are consistent with the corresponding PDG value, $m_{B^0} = 5279\pm 0.20  ~MeV $   \cite{PDG} and the Flavour Lattice Averaging Group (FLAG) value $f_{B^0} = 190.0(1.3) ~MeV$ \cite{Flag},  with the deviations remaining below $1 \%$.
\item \textbf{Medium modified QCD condensates:} In the next stage of the analysis, our primary objective is to investigate how the quantities under consideration evolve with temperature and chemical potential in the medium. Therefore, expressions describing the dependence of quantities such as the quark condensate, gluon condensate, continuum threshold, strong coupling constant, and energy–momentum tensor on both temperature and chemical potential are required. Although these quantities have been extensively studied at zero chemical potential and finite temperature, studies addressing simultaneously finite values of both variables remain limited. Their careful specification is consequently necessary for a reliable analysis. To this end, we use the temperature and chemical potential dependent quark condensate, $ \langle  \bar{q}q \rangle_{(T,\mu)} $, as follows
\begin{eqnarray}\label{quarkkondensat}
	\langle  \bar{q}q \rangle_{(T,\mu)}=-\frac{N_c}{4\pi^2}\Big (0.1842 -0.1669 \mu^2  - 5.6272\mu^2 T +1.9337 T-13.1064 T^2\Big )~~.
\end{eqnarray}
The above functional form of the quark condensate was obtained by fitting the data presented in Ref. \cite{Colangelo}. The medium-dependent gluon condensate in terms of the vacuum gluon condensate and the quark condensate can be expressed  \cite{Gubler} as

\begin{eqnarray}\label{G2TLattice}
	\delta \langle \frac{\alpha_{s}}{\pi}G^{2}\rangle_{(T,\mu)}&=&\langle  \frac{\alpha_s}{\pi} G^2  \rangle_{0,0}-\frac{8}{9}[ \delta T^{\mu}_{\mu}(T)-m_{u} \delta \langle\bar{u}u\rangle_{(T,\mu)}-m_{d} \delta \langle\bar{d}d\rangle_{(T,\mu)}-m_{s} \delta \langle\bar{s}s\rangle_{(T,\mu)}], 
\end{eqnarray}
here the medium-modified strange quark condensate defined by $ \langle  \bar{s}s \rangle_{(T,\mu)}=0.8\langle  \bar{q}q \rangle_{(T,\mu)}$. The quantity $f(T)$ is characterized by subtracting the corresponding vacuum values as $ \delta f(T)\equiv f(T)-f(0) $. Accordingly, the trace anomaly is defined by $ \delta T^{\mu}_{\mu}(T)=\varepsilon(T)-3p(T) $, where $\varepsilon(T)$ and $ p(T) $ represent the energy density and pressure, respectively. To describe the temperature dependence of $ \delta T^{\mu}_{\mu}(T) $, we employ the parametrization obtained by fitting the lattice QCD results of Ref.~\cite{Bazavov, Borsanyi}, as carried out in our previous work \cite{AziziTurkan}. The resulting parametrization is written as
\begin{eqnarray}\label{epsmines3p}
	\delta T^{\mu}_{\mu}(T) &=& T^{4} \Big (0.115+ 0.020 exp \Big [\frac{T}{0.034[GeV]}\Big ] \Big ).
\end{eqnarray}
 Another quantity required for our numerical analysis is the in medium energy density. In the present calculations, only its gluonic contribution enters the formalism. We employ the parametrization obtained in our previous work by fitting the data reported in Ref.~\cite{Bazavov}, which is given below
 \begin{eqnarray}\label{tetag}
 	\langle\Theta^{g}\rangle&=&  0.091 exp[\frac{T}{0.047[GeV]}]T^{4}-0.731T^{4}.
 \end{eqnarray}
 The analysis requires the temperature-dependent strong coupling $g_s^2(T)$, and we use the following parametrization introduced in Ref.~\cite{Kaczmarek,Morita}:
 \begin{eqnarray}\label{geks2T}
 	g_s^{-2}(T)=\frac{11}{8\pi^2}\ln\Big(\frac{2\pi
 		T}{\Lambda_{\overline{MS}}}\Big)+\frac{51}{88\pi^2}\ln\Big[2\ln\Big(\frac{2\pi
 		T}{\Lambda_{\overline{MS}}}\Big)\Big],
 \end{eqnarray}
 where, $\Lambda_{\overline{MS}}\simeq T_{c}/1.14$ and $T_c$ denotes the critical temperature. Although the term critical temperature is commonly used in discussions of the transition to the quark-gluon plasma (QGP), no unique and universally established value can be assigned to it. The QCD phase diagram is expected to exhibit different transition regimes, including a crossover region, a critical end point (CEP), and a first-order phase transition line. At vanishing or relatively small chemical potential, lattice QCD studies characterize the transition in terms of a pseudo-critical temperature, $T_{pc}$, and indicate a smooth crossover from hadronic matter to the QGP around $T_{pc}\backsimeq155~MeV$ \cite{Bhattacharya, Bazavov3}. As the chemical potential increases, the nature of the transition may change from a crossover to a first-order transition, with the crossover boundary potentially terminating at a CEP. A transition associated with the formation of the QGP, however, may require considerably higher temperatures, depending on the region of the $T - \mu $ phase diagram being considered. Based on the variation of the critical temperature with chemical potential reported in Ref.~\cite{Colangelo}, we obtain the following fitting function
 \begin{eqnarray}\label{criticaltemp}
 T_c(\mu) = 0.21(1.00-1.54\mu^2-0.68\mu^4).
  \end{eqnarray}
  Here $T_c = 0.21 ~GeV$ in the limit of $\mu = 0$. Finally, the in-medium continuum threshold $s_{0}(T,\mu)$ appearing in Eq.~(\ref{sumrule}) is given by \cite{Ayala1}
 \begin{eqnarray}\label{Conthreshold}
 	s_{0}(T,\mu) \backsimeq s_{0}(0,0) \Big [\frac{\langle  \bar{q}q \rangle_{(T,\mu)}}{\langle \bar{q}q \rangle_{0,0}} -\frac{T^{2}/3-\mu^{2}/\pi^2}{2f_{\pi}^2(0,0)}\Big ],
 \end{eqnarray}
 here $ f_{\pi}(0,0)=92.21\pm0.14~MeV $ \cite{Nakamura}.
 
 \item \textbf{Behavior under temperature and chemical potential variations:} To examine the behavior of the mass and decay constant with respect to temperature and chemical potential, we present their plots in Figs.  (\ref{massTmu}) and (\ref{decayTMu}). These graphs offer a valuable understanding of their in-medium effects, where the observed behavior of the mass and decay constant, both directly tied to the order parameter of chiral symmetry breaking,  provides insight into the restoration of chiral symmetry within the medium. 
 
  In Fig.  (\ref{massTmu}), the left hand panel illustrates the temperature dependence of the mass for several fixed values of $ \mu$. The graph clearly shows that, for all $\mu$ values considered, the mass exhibits a slight initial increase with temperature before decreasing toward zero. A closer inspection of the curves reveals that, at low temperatures and low chemical potentials, the mass increases by approximately $5\%$ up to $T=60~\mathrm{MeV}$ for all $\mu$ values except $\mu=0.4~\mathrm{GeV}$, for which the increase is about $3\%$ at $T=40~\mathrm{MeV}$. Thus, the mass may be regarded as remaining nearly constant in the low-temperature region rather than exhibiting a significant increase. This conclusion is more clearly illustrated in the right panel. In the left panel, variations in the chemical potential do not affect the magnitude of the mass increase up to $T=20~\mathrm{MeV}$. The influence of the chemical potential on the decrease in mass becomes noticeable at different temperatures. Moreover, the temperature at which the mass vanishes shifts to lower values as $\mu$ increases. For $\mu=0$, the mass reaches zero at approximately $T=273~\mathrm{MeV}$, whereas for $\mu=0.4~\mathrm{GeV}$, this temperature decreases to approximately $182~\mathrm{MeV}$. This trend is consistent with the expected behavior associated with phase transitions in the QCD phase diagram.
 
 On the right panel of Fig. (\ref{massTmu}), the change in mass as a function of chemical potential is plotted for specific temperature values. Across all temperature values, the curves indicate that the mass decreases as the chemical potential increases. As previously highlighted in the mass–temperature graph, the chemical potential value at which the mass vanishes shifts toward lower values as the temperature increases. At zero temperature, the decrease in mass is slower and the zero point is reached at very large $\mu$ values. In contrast, in the temperature range $T=50–100 ~MeV $ the mass reduction becomes more moderate and occurs within physically relevant ranges of the chemical potential. Finally, at $T=155 ~MeV$, the mass reaches zero at $\mu=0.38 ~GeV$, accompanied by a noticeably sharp decrease.
 \begin{figure}[!h]
 	\centering
 	\begin{minipage}{0.48\textwidth}
 		\centering
 		\includegraphics[width=\linewidth]{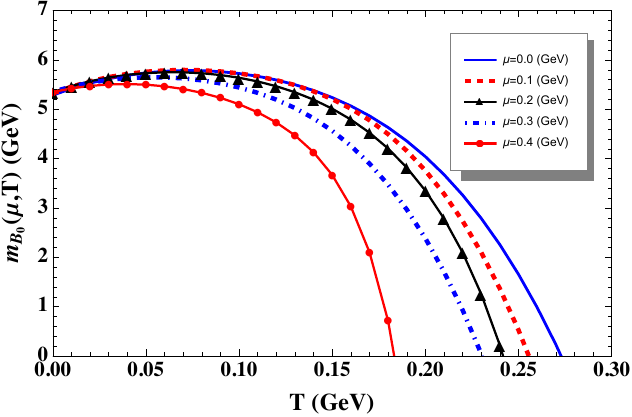}
 		\label{fig:left}
 	\end{minipage}
 	\hfill 
 	\begin{minipage}{0.48\textwidth}
 		\centering
 		\includegraphics[width=\linewidth]{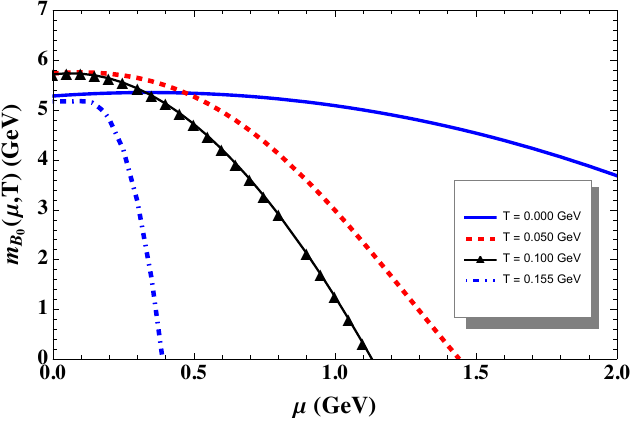}
 		\label{fig:right}
 	\end{minipage}
 	\vspace{-10pt}
 	\caption{Mass variations of $B^{0}$ meson with respect to chemical potential (right panel) and temperature (left panel).}
 	\label{massTmu}
 \end{figure}

 Furthermore, to determine whether any particular QCD condensate dominates the observed change in the mass, the percentage contributions of the perturbative term and QCD condensate terms are calculated at fixed temperature as a function of the chemical potential. The corresponding results are presented in Table \ref{tab:massyuzde}. As can be seen, no single condensate provides a dominant contribution; rather, their contributions to the mass variation are of comparable magnitude. Note that this table shows percentage values of different condensates contributions on the variation of mass and is different than the OPE convergence of the sum rules discussed above. The shown value for each temperature is obtained by taking the averages of the results obtained by   varying $ \mu $ in the considered interval. 
 \begin{table}[ht!] 
 	\centering
 	\begin{tabular}	{ |c|c|c|c|c|} 
 		\hline \hline
 		Temperature  & Perturbative  & $ \langle  \bar{d}d \rangle_{T,\mu} $  & $ {\langle}  \frac{\alpha_s}{\pi} G^2 {\rangle}_{T,\mu}$ & $\langle \bar{q}g_{s}\sigma_{\mu \nu} G^{\mu \nu}q \rangle_{T,\mu}$    \\ \hline
 		$T=0 ~MeV$  &  $ 25.2\% $   & $24.9\%$& $25.0\%$ & $24.9\%$ \\   \hline
 		$T=50 ~MeV$  & $ 25.0\% $  & $ 25.0\% $ & $ 25.0\% $& $25.0\%$  \\    \hline
 		$T=100 ~MeV$     & $ 26.0\% $  &$ 24.6\% $ &$ 24.7\% $ & $24.7\%$ \\  \hline
 		$T=155~MeV$       &  $ 25.3\% $   & $ 24.6\% $&$ 25.1\% $& $25.0\%$  \\  \hline
 		\hline \hline
 	\end{tabular}
 	\caption{Relative percentage contributions of perturbative and non-perturbative QCD condensates to mass shifts. The reported values for different temperatures are obtained by taking the averages of the results obtained by   varying $ \mu $ in the considered interval. }
 	\label{tab:massyuzde}
 \end{table}

 In addition to the mass, investigating the in-medium behavior of the decay constant is equally important, since it is closely related to the quark condensates. Moreover, its variation under extreme conditions is of particular significance, as the decay constant is directly required as an order parameter for the restoration of chiral symmetry. Fig.  (\ref{decayTMu}) was drawn in this context. When the left panel of Fig.  (\ref{decayTMu}), which displays the temperature dependence of the decay constant, is examined, the decay constant is seen to increase initially with temperature. It then starts to decrease and eventually vanishes.  Similar to the behavior observed for the mass, the temperature thresholds at which the decay constant drops to zero shift to lower values as the chemical potential increases. Notably, unlike the mass, the data obtained for $\mu=0.0-0.1 ~GeV $remain practically insensitive to the considered temperature range, whereas the effect of increasing temperature and chemical potential on the decrease of the decay constant becomes more pronounced at larger values of $\mu$. In particular, the influence of temperature on the decay constant becomes clearly noticeable above $\mu=0.1 ~GeV$. While for $\mu=0.4 ~GeV $ the temperature effect sets in immediately,  within the range of  $\mu=0.0-0.4 ~GeV $ the effect starts to emerge at a relatively low temperature of about $T=14 ~MeV$. The temperature range at which the decay constant drops to zero is $T=173 ~MeV$ for $\mu=0.1 ~GeV$ and $T=118 ~MeV$ for $\mu=0.4 ~GeV$.

 The right panel of Fig.  (\ref{decayTMu}) presents the dependence of the decay constant on the chemical potential at selected temperature values, allowing the effect of temperature on the decay constant to be seen more clearly. In accordance with the behavior described above, the chemical potential at which the decay constant vanishes shifts toward lower values as the temperature increases. This range decreases from $\mu=0.83 ~GeV$ at $T=0.000 ~MeV$ to $\mu=0.23 ~GeV$ at $T=155 ~MeV$.
 \begin{figure}[!h]
 	\centering
 	\begin{minipage}{0.48\textwidth}
 		\centering
 		\includegraphics[width=\linewidth]{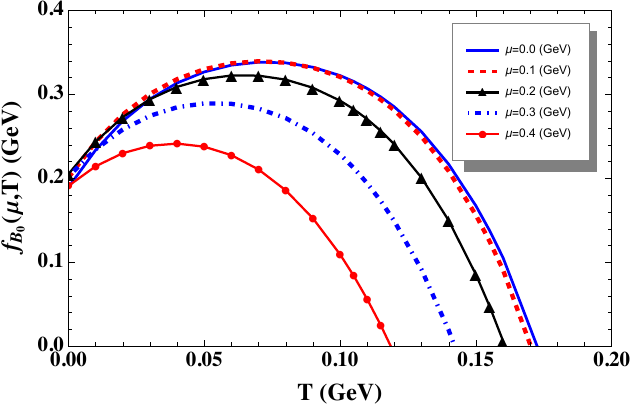}
 		\label{fig:left}
 	\end{minipage}
 	\hfill 
 	\begin{minipage}{0.48\textwidth}
 		\centering
 		\includegraphics[width=\linewidth]{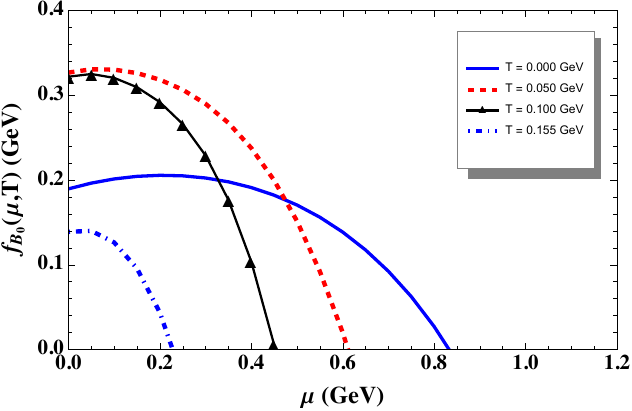}
 		\label{fig:right}
 	\end{minipage}
 	\vspace{-10pt}
 	\caption{The behavior of 
 		 decay constant as a function of temperature (left panel) and chemical potential (right panel).}
 	\label{decayTMu}
 \end{figure}	

 The percentage contributions of the condensates to the decay constant's variations is presented in Table \ref{tab:decayyuzde}. Although the percentage contribution of the quark condensate increases with temperature, the magnitude of this increase remains insignificant. Consequently, no dominant contribution is observed with variations in temperature and chemical potential. Their effects on the change in the decay constant are comparable.
 \begin{table}[ht!] 
 	\centering
 	\begin{tabular}	{ |c|c|c|c|c|} 
 		\hline \hline
 		Temperature  & Perturbative  & $ \langle  \bar{d}d \rangle_{T,\mu} $  & $ {\langle}  \frac{\alpha_s}{\pi} G^2 {\rangle}_{T,\mu}$ & $\langle \bar{q}g_{s}\sigma_{\mu \nu} G^{\mu \nu}q \rangle_{T,\mu}$    \\ \hline
 		$T=0 ~MeV$  &  $ 25.8\% $   & $24.8\%$& $24.7\%$ & $24.7\%$ \\   \hline
 		$T=50 ~MeV$  & $ 23.5\% $  & $ 25.5\% $ & $ 25.5\% $& $25.5\%$  \\    \hline
 		$T=100 ~MeV$     & $ 21.4\% $  &$ 26.3\% $ &$ 26.2\% $ & $26.1\%$ \\  \hline
 		$T=155~MeV$       &  $ 23.6\% $   & $ 27.1\% $&$ 24.7\% $& $24.6\%$  \\  \hline
 		\hline \hline
 	\end{tabular}
 	\caption{Percentage contributions of perturbative and QCD condensate terms to decay constant variations. The reported values for different temperatures are obtained by taking the averages of the results obtained by   varying $ \mu $ in the considered interval. }
 	\label{tab:decayyuzde}
 \end{table}
\end{itemize}

\section{Summary and Concluding remarks}  \label{sec:3}

In this work, we have systematically investigated the in-medium properties of the  $B^0 $ meson within the framework of QCD two point sum rules, with particular emphasis on the effects of temperature ($T$) and chemical potential ($\mu $). By introducing the temperature and chemical potential dependence of the relevant QCD condensates into the OPE, including the quark condensate ($\langle \bar{q}q\rangle_{T,\mu}$), the associated mixed condensate ($\langle\bar{q}g_s\sigma_{\mu \nu} G^{\mu \nu}q\rangle_{T,\mu}$) and gluon condensate ($\langle\frac{\alpha_s}{\pi}G^2\rangle_{T,\mu}$), we obtained sum rules that establish a consistent connection between the underlying quark–gluon dynamics and the hadronic observables under consideration.

To determine the consistency of the QCD sum rules in a hot and dense medium for the investigated quantities, the working windows of the Borel parameter and the continuum threshold were established at zero temperature. The resulting plot given in Fig. (\ref{vacuumMassG}) indicates almost complete insensitivity to these parameters. Furthermore, all input parametrizations employed in numerical analysis reproduce mass and decay constant values consistent with the PDG data in the limits $T\rightarrow 0$ and $\mu \rightarrow 0$.

In the subsequent analysis, as the primary objective of this study, the temperature and chemical potential dependence of the mass and decay constant of the $B^0$ meson was investigated through the corresponding plots. The data presented in these plots require knowledge of the functional dependence of the medium-modified condensates and other relevant quantities. Since explicit expressions describing the dependence of these quantities on both temperature and chemical potential are unavailable in the literature, a suitable reformulation of these expressions is required. For this purpose, the expressions for the gluon condensate, continuum threshold, and strong coupling are adopted from the literature. However, the quark condensate appearing in these expressions, which depends on both temperature and chemical potential, as well as the chemical potential dependence of the critical temperature, are obtained by fitting the corresponding results reported in the relevant studies. 

As can be seen from the mass and decay constant plots, the chemical potential values at which the mass and decay constant vanish decrease with increasing temperature. For the mass, the corresponding chemical potential ranges from $\mu=0.38 ~GeV$ to $\mu=1.43 ~GeV$, whereas for the decay constant, it ranges from $\mu=(0.23-0.83)~GeV$. This trend is consistent with the behavior depicted in the QCD phase diagram in the ($T-\mu$) plane. However, these numerical values are not intended to claim a precise determination of the critical point, but rather to indicate the range of temperature and chemical potential over which the particle under consideration approaches the chiral transition from the hadronic phase under the influence of the medium. At the pseudo-critical temperature, $T_{pc}=155~ MeV$, the mass exhibits a decrease of approximately $ 4\% $  for $\mu=(0.0-0.2) ~GeV$, $ 17\% $  for $\mu=0.3 ~GeV$, and $ 38\% $  for $\mu=0.4 ~GeV$. The corresponding decrease in the decay constant is approximately $ 37.5\% $  for $\mu=(0.0-0.1) ~GeV$ and $ 76\% $  for $\mu=0.2 ~GeV$, whereas for larger values of $\mu$, it vanishes before reaching  $T_{pc}$. It has been observed that the effect of temperature  becomes more dominant with increasing the chemical potential, particularly for the decay constant. With variations in temperature or chemical potential, the changes in the quantities of the particle under consideration are found to receive comparable contributions from the perturbative and condensate terms. Regarding the behavior observed at low temperatures, the initial increase in the $B^0$ meson mass may be related to the adopted quark condensate parameterization, since the absolute value of the quark condensate exhibits a similar temperature dependence, with both quantities reaching their maximum values at approximately the same temperature. However, the observed behavior reflects the combined effects of the perturbative and non-perturbative contributions entering the sum rules and cannot be attributed solely to the quark condensate.

The results obtained in this study are expected to contribute to the interpretation and generation of data from heavy-ion collision experiments that investigate $B$-mesons or QGP signatures, such as ALICE, LHCb at CERN and STAR, PHENIX at RHIC, as well as from experiments aimed at exploring dense-matter environments, such as FAIR-CBM, NICA, and J-PARC.



\section*{DECLARATION OF AI-ASSISTED LANGUAGE EDITING}

The authors used artificial intelligence (AI) tools solely for language editing and improving the readability of the manuscript. All scientific content, including the analytic and numerical computations, interpretation of the results and the conclusions, was developed and verified by the authors, who take full responsibility for the scientific content of this work.

\end{document}